\documentclass[aps,prl,nofootinbib,twocolumn,floatfix]{revtex4-2}

\usepackage{amsmath,amssymb,bm,graphicx,booktabs,slashed}
\usepackage{xcolor,needspace}
\usepackage[colorlinks=true,citecolor=blue,linkcolor=blue,urlcolor=blue]{hyperref}
\newcommand{\MSbar}{\overline{\rm MS}}
\newcommand{\as}{\alpha_s}
\newcommand{\Qtop}{\mathcal Q_{\rm top}}

\newcommand{\dd}{\mathrm d}

\begin{document}

\title{ALP Distributions from Hadron Structure}

\author{Shuai Zhao}
\email{zhaos@tju.edu.cn}
\affiliation{Department of Physics and Center for Joint Quantum Studies,
 School of Science, Tianjin University, Tianjin 300350, China}

\date{\today}

\begin{abstract}
We investigate how hadron structure and quark-flavor
couplings shape axion-like particle (ALP) radiation
in hadron collisions. For the elastic pole, the virtuality integral is ultraviolet
finite at fixed momentum fraction if the pseudoscalar form
factor falls as a positive power of $1/Q^2$ at large $Q^2$,
eliminating the point-particle beam-energy logarithm. Pion-pole and lattice-QCD isovector benchmarks suppress the process-weighted elastic contribution by roughly three orders of magnitude relative to the point-particle result. Two coefficient sets matched to the same leading low-energy
nucleon source yield process-weighted resolved leading
logarithms differing by four orders of magnitude,
demonstrating the sensitivity of resolved ALP radiation
to heavy-quark coupling configurations.
\end{abstract}

\maketitle

\emph{Introduction.---}
High-energy hadron beams can act as sources of quasireal particles,
effectively turning a hadron collider into a collider of the fields
they radiate
\cite{vonWeizsacker:1934nji,Williams:1934ad,Budnev:1975poe}.
This framework also applies to weakly coupled pseudoscalars.
Axion-like particles (ALPs) extend the QCD-axion paradigm to
independent masses and couplings, thereby admitting flavor structures
that can be probed at high-energy colliders
\cite{Peccei:1977hh,Peccei:1977ur,Weinberg:1977ma,
	Wilczek:1977pj,Masso:1995tw,Jaeckel:2010ni}.
A recent proposal extends this equivalent-particle picture to ALPs,
opening ALP--photon collisions in ultraperipheral $p$Pb events as
probes of flavor couplings~\cite{Barbosa:2025zyn}.
For a composite emitter, however, the putative ALP flux is not a
single universal object. Intact-hadron, dissociative, and inclusive
selections probe distinct components of the hadronic response.

This distinction matters because the ALP--nucleon coupling
at zero momentum transfer, $g_{aNN}(0)$, fixes the normalization
of the elastic pole, whereas its form factor, the dissociative
spectrum, and resolved radiation contain independent hadronic
and short-distance information.  This separation
parallels that in the photon parton distribution function (PDF) of the proton, for which elastic and inelastic
contributions form a common renormalized distribution only after
factorization and matching
\cite{Manohar:2016nzj,Manohar:2017eqh}.  An ALP distribution must
likewise be specified together with the selected hadronic final state
and the factorization scheme.

In this Letter we formulate this separation at $O(C^2)$ in the ALP source coefficients.  We first prove that any positive-power falloff of the elastic
form factor removes the point-particle beam-energy logarithm and
quantify the resulting saturation using pion-pole and lattice-QCD
benchmarks. We then derive the resolved-quark leading logarithm,
exhibit two short-distance coefficient sets that match to the same
leading low-energy source but yield resolved contributions differing
by four orders of magnitude, and identify the matching structure
required for an inclusive distribution.

\emph{ALP distribution.---}
Following the scalar-PDF definition of
Ref.~\cite{Fornal:2018znf}, we define the ALP distribution
in the $\MSbar$ scheme by
\begin{align}
 f_{a/p}^{\MSbar}(x,\mu)= {xP^+}\int \frac{\dd\lambda}{2\pi}
 e^{-ixP^+\lambda}
 \langle P|a(\lambda \bar n)a(0)|P\rangle_{\MSbar}.
 \label{eq:pdfdef}
\end{align}
We use $v^+=\bar n\cdot v$, $\bar n^2=0$, and $P^+>0$, where the prefactor $xP^+$ gives the ALP number-density normalization. The proton-spin is averaged.  The equation of motion, $(\Box+m_a^2)a=J_a$ with $J_a=f_a^{-1}C_IO_I
$,
relates the connected bilocal to the source spectral density $W_{IJ}$ (End Matter). Its one-proton pole involves $\langle p(P')|O_I|p(P)\rangle={\cal F}_I(Q^2)\bar u(P')i\gamma_5u(P)$, while all other states describe dissociation. We denote the positive spectral boundary restricted to $Q^2<\mu_0^2$ by
\begin{equation}
 {\cal B}_{a/p}^{\rm cut}(x,\mu_0)={\cal B}_{a/p}^{\rm el,<}(x,\mu_0)
 +{\cal B}_{a/p}^{\rm diss,<}(x,\mu_0).
 \label{eq:lowboundary}
\end{equation}
Physically, ${\cal B}_{a/p}^{\rm cut}(x,\mu_0)$ is the
cutoff-defined low-virtuality contribution to the ALP number density,
obtained by summing over intact-proton and dissociative hadronic final
states with $Q^2<\mu_0^2$.
This cutoff quantity is distinct from the renormalized PDF $f_{a/p}^{\MSbar}(x,\mu_0)$, which also contains a scheme-conversion term specified in the End Matter. 

\emph{Ultraviolet saturation of the elastic pole.---}
The recoil kinematics sets the minimum virtuality sampled
by the elastic form factor. For a hadronic recoil state
$X$ of mass $M_X$ in $p(P)\to X(P_X)+a^*(q)$, define
$x=q^+/P^+$ and $Q^2=-q^2$, then the on-shell conditions give
\begin{align}
	Q^2&=\frac{q_T^2+x^2M^2+x(M_X^2-M^2)}{1-x},
	\nonumber\\
	Q_{\min}^2&=\frac{x[M_X^2-(1-x)M^2]}{1-x},
	\label{eq:kin}
\end{align}
where the minimum occurs at $q_T=0$.
The elastic pole corresponds to $X=p$, for which
$Q_{\min}^2=x^2M^2/(1-x)$. Writing the elastic source vertex as
$g_{aNN}(Q^2)\bar u i\gamma_5u$,
we obtain the virtuality density
\begin{align}
 {\cal I}^{\rm el}(x,Q^2)
 =\frac{x}{16\pi^2}\frac{Q^2|g_{aNN}(Q^2)|^2}{(Q^2+m_a^2)^2},
 \label{eq:elastic}
\end{align}
where the spin-averaged pseudoscalar trace supplies the numerator $Q^2$.  At finite beam energy, the elastic distribution at fixed $0<x<1$ is
\begin{align}
 f^{\rm el}_{a/p}(x,E_p)=
 \int_{Q_{\min}^2(x)}^{Q_{\rm up}^2(x,E_p)}\dd Q^2\,
 {\cal I}^{\rm el}(x,Q^2),
 \label{eq:elasticfinite}
\end{align}
where $Q_{\rm up}^2(x,E_p)$ is the kinematic upper limit and tends to infinity as $E_p\to\infty$ at fixed $x$.  For $m_a=0$, the general saturation condition is
\begin{align}
 \int_{\Lambda_{\rm had}^2}^{\infty}\frac{\dd Q^2}{Q^2}
 |g_{aNN}(Q^2)|^2<\infty,
 \label{eq:integrability}
\end{align}
where $\Lambda_{\rm had}$ denotes a fixed hadronic scale.
The convergence condition in Eq.~\eqref{eq:integrability}
ensures that Eq.~\eqref{eq:elasticfinite} saturates as
$E_p\to\infty$ at fixed $x$, eliminating the beam-energy
logarithm. For $|g_{aNN}(Q^2)|\le K(\Lambda_{\rm had}^2/Q^2)^p$
at large virtuality with $p>0$, the approach to saturation
is controlled by
\begin{align}
\Delta f^{\rm el}_{a/p}
\le \frac{xK^2}{32\pi^2p}
\left(\frac{\Lambda_{\rm had}^2}{Q_{\rm up}^2}\right)^{2p},
\end{align}
provided $Q_{\rm up}^2$ lies in this asymptotic region.
This fixed-$x$ statement does not exclude energy dependence generated after convolution over $x$, for example from a moving small-$x$ region or the hard luminosity.  QCD form-factor power counting satisfies the sufficient condition \cite{Brodsky:1973kr,Lepage:1980fj}.

For comparison, a constant point-nucleon coupling $g_p$
gives a beam-energy logarithm when integrated up to
$q_{T,\max}\sim xE_p$:
$f^{\rm pt}_{a/p}\simeq
g_p^2x\ln(E_p^2/M^2)/(16\pi^2)$ for $m_a=0$.
Alternatively, imposing a virtuality cutoff $Q<M$
replaces the energy-dependent upper limit by $M^2$
at sufficiently high beam energy, yielding
\begin{align}
 f^{Q<M}_{a/p}(x)&=\frac{g_p^2x}{16\pi^2}\ln\frac{1-x}{x^2}
 \Theta(x_{\max}-x),
 \label{eq:sharp}
\end{align}
The endpoint $x_{\max}=\frac{\sqrt5-1}{2}$ follows from
$Q_{\min}^2(x_{\max})=M^2$.
While Ref.~\cite{Barbosa:2025zyn} uses this condition
to restrict $x$ after the pointlike virtuality integration,
Eq.~\eqref{eq:sharp} implements the cutoff $Q^2<M^2$
within the integral itself.

To quantify the additional suppression from the nucleon form factor, we first consider the isovector source, whose low-virtuality behavior is constrained by chiral symmetry. In the isospin limit, define $A_3^\mu=(\bar u\gamma^\mu\gamma_5u-\bar d\gamma^\mu\gamma_5d)/2$, then $O_3=\hat m(\bar ui\gamma_5u-\bar di\gamma_5d)=\partial_\mu A_3^\mu$, and $\langle0|A_3^\mu|\pi^0(q)\rangle=if_\pi q^\mu$ fixes
\begin{align}
 \langle X|O_3|p\rangle=
 \frac{f_\pi m_\pi^2}{m_\pi^2+Q^2}\Gamma_X^{(\pi)}+R_X(Q^2),
 \label{eq:ward}
\end{align}
where $\Gamma_X^{(\pi)}$ denotes the residue fixed by the on-shell $\pi p\to X$ amplitude, analytically continued to the pion pole in a common phase convention, and $R_X$ is regular there.  For elastic proton emission, the axial Ward identity fixes $g_{aNN}^{(3)}(0)=C_3g_AM/f_a$. The pion-pole approximation gives the normalized coupling $F_3(Q^2)\equiv g_{aNN}^{(3)}(Q^2)/g_{aNN}^{(3)}(0)\simeq m_\pi^2/(m_\pi^2+Q^2)$~\cite{Bishara:2017pfq}. For $m_a=0$, inserting this approximation into Eq.~\eqref{eq:elastic} and integrating to infinite virtuality yields
\begin{align}
 f_{a/p}^{\rm el,\pi}(x)&=\frac{|g_{aNN}^{(3)}(0)|^2x}{16\pi^2}
 \int_{Q_{\min}^2(x)}^\infty\frac{\dd Q^2}{Q^2}
 \left(\frac{m_\pi^2}{m_\pi^2+Q^2}\right)^2\nonumber\\
 &=\frac{|g_{aNN}^{(3)}(0)|^2x}{16\pi^2}
 \left[\ln\!\left(1+\frac{m_\pi^2}{Q_{\min}^2}\right)
 -\frac{m_\pi^2}{m_\pi^2+Q_{\min}^2}\right]\nonumber\\
 &\xrightarrow[]{Q_{\min}^2\gg m_\pi^2}
 \frac{|g_{aNN}^{(3)}(0)|^2m_\pi^4(1-x)^2}{32\pi^2x^3M^4}.
 \label{eq:closed}
\end{align}
This strong suppression is realized by the lattice pseudoscalar-density form factor $G_5$ \cite{Alexandrou:2023qbg}. We use its physical-point continuum $z^3$ fit, normalize at $Q^2=0$, truncate at the fit limit $Q^2=1~{\rm GeV}^2$, and propagate the published covariance. Diagnostic continuations above that limit are discussed separately in the End Matter. 

For a general ALP, light-quark isoscalar and gluonic sources may also contribute, and the elastic coupling becomes
\begin{align}
 g_{aNN}(Q^2)=g_3F_3(Q^2)+g_0F_0(Q^2)+g_GF_G(Q^2).
 \label{eq:sourcebasis}
\end{align}
Here $g_i=C_i{\cal F}_i(0)/f_a$ is the zero-virtuality coupling of each source and $F_i(Q^2)={\cal F}_i(Q^2)/{\cal F}_i(0)$, so that $F_i(0)=1$ and $g_3=g_{aNN}^{(3)}(0)$. The additional sources are $O_0=\hat m(\bar ui\gamma_5u+\bar di\gamma_5d)$ ($I=0$) and $O_G=\Qtop=(\as/8\pi)G\widetilde G$. Their amplitudes interfere with the isovector term. The illustrative $F_0$ and $F_G$ profiles are given only in the End Matter.  A complete flavor decomposition requires disconnected and singlet matrix elements \cite{Green:2017keo,Alexandrou:2021wzv,Barone:2025rye,Barone:2026uyx,Fukushima:2026sot}, while QCD-axion couplings additionally require model-dependent matching \cite{GrilliDiCortona:2015jxo}.

For a spin-zero nucleus, the local pseudoscalar source
is further constrained by parity. In parity-conserving
strong interactions,
$\langle A(0^+,P')|\mathcal P(0)|A(0^+,P)\rangle=0$,
since the two external momenta cannot form a pseudoscalar. Here $\mathcal P$ denotes any local pseudoscalar QCD source,
such as $O_3$, $O_0$, $O_G$, or their linear combination.
Thus $^{208}{\rm Pb}(0^+)$ supplies no coherent elastic
ALP source of this form. Spinful targets, parity-changing
transitions, and breakup involve different matrix elements.

Beyond the elastic pole, a heavier hadronic recoil state
raises the minimum virtuality by $x(M_X^2-M^2)/(1-x)$.
A shifted pion-pole kernel illustrates this kinematic
suppression in the End Matter. The dissociative rate
further depends on transition spin structures, spectral
strength, and the regular terms in Eq.~\eqref{eq:ward}.

\emph{Resolved radiation and matching requirements.---}
The elastic pole and dissociative spectrum supply the low-virtuality
contributions to the ALP distribution, whereas an inclusive distribution
at a hard scale also contains perturbatively resolved radiation. We now
derive its quark-initiated leading-logarithmic (LL) evolution, identify the boundary conversion
and hard subtraction required for consistent matching, and use
heavy-flavor thresholds to show that the same low-energy source can yield
different resolved components.

At partonic scales, we describe the ALP interaction by the
renormalized quark and gluon pseudoscalar sources
\cite{Bauer:2017ris,Bauer:2020jbp},
\begin{align}
	J_a(\mu)=\frac1{f_a}\left[
	\sum_q C_q(\mu)m_q(\mu)\bar q i\gamma_5 q
	+\widetilde C_G(\mu)\Qtop\right].
	\label{eq:qcdsource}
\end{align}
Here
$\Qtop=(\as/8\pi)G\widetilde G
=(\as/16\pi)G_{\mu\nu}^a
\epsilon^{\mu\nu\rho\sigma}G_{\rho\sigma}^a$,
$q$ runs over the active flavors, and all masses, operators, and
Wilson coefficients are defined at the common renormalization and
factorization scale $\mu$.  The parameter $f_a$ denotes the ALP
interaction scale.

Collinear emission from the effective coupling $C_qm_q/f_a$ gives
$\dd f_{a/q}/(\dd z\,\dd k_T^2)=|C_qm_q(k_T)/f_a|^2z/(16\pi^2k_T^2)$, where $z$ is the fraction of the parent-quark momentum carried by the ALP. The splitting kernel is therefore proportional to $z$ and has no soft $1/z$ enhancement~\cite{Chen:2016wkt}. At $O(C_q^2)$, retaining QCD evolution of the emitting quark, the LL evolution from $\mu_0$ to $\mu$ is
\begin{align}
 f_{a/p}^{\MSbar}&(x,\mu)-f_{a/p}^{\MSbar}(x,\mu_0)
 =\frac{x}{16\pi^2f_a^2}\sum_q|C_q|^2\nonumber\\
 &\times\int_{\mu_q^2}^{\mu^2}\frac{\dd k^2}{k^2}m_q^2(k)
 \int_x^1\frac{\dd\xi}{\xi^2}f_{q+\bar q/p}(\xi,k)+\cdots.
 \label{eq:matchpdf}
\end{align}
Here $\xi$ is the parent-quark momentum fraction, $k$ is the ordered collinear scale, and $f_{q+\bar q/p}$ is the quark-plus-antiquark proton PDF. We take $\mu_q=\max(\mu_0,m_q^{\rm thr})$, where $m_q^{\rm thr}$ is the flavor-activation threshold in the chosen PDF scheme. The ellipsis denotes terms beyond LL evolution. The Wilson coefficients are held fixed at this accuracy, while the masses run. For a generic flavor $q$, Eq.~\eqref{eq:matchpdf} retains the quark PDF at the emission scale $k$. For heavy flavors $Q=c,b$, the scale dependence generated
by $g\to Q\bar Q$ contributes at LL accuracy.
Evaluating $f_{Q+\bar Q/p}$ at the fixed upper scale $\mu$
rather than the emission scale $k$ changes the coefficient
of the leading $\as\ln^2(\mu^2/m_Q^2)$ term. The same ordering must be implemented in the
hard-coefficient subtraction of
Eq.~\eqref{eq:subtraction}.

The gluonic source has different collinear power counting. For
on-shell external gluons with momentum transfer $r=p-p'$, its squared
matrix element is proportional to $(p\cdot p')^2=Q^4/4$. It therefore
cancels the squared ALP propagator as $Q^2\to0$ and produces no
real-emission collinear logarithm at this order. Gluon-initiated
processes nevertheless contribute through a finite, process-dependent
hard coefficient. With the
normalization in Eq.~\eqref{eq:qcdsource}, QCD mixing of the gluonic
source into the fermion coefficients starts at $O(\as^2)$
\cite{Bauer:2020jbp}.
For the numerical comparison, we retain the quark LL term
in Eq.~\eqref{eq:matchpdf}, leaving the finite
cutoff-to-$\MSbar$ boundary conversion and matched hard
terms, including the gluon-initiated coefficient, unevaluated.
The quoted PDF and scale variations refer to this LL component.

\begin{figure*}[ht]
\centering
\begin{minipage}[t]{0.478\textwidth}
\centering
\includegraphics[width=\linewidth]{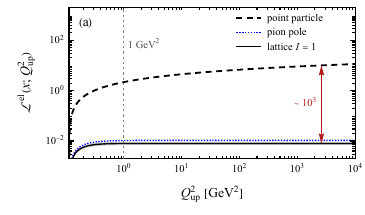}
\end{minipage}\hfill
\begin{minipage}[t]{0.478\textwidth}
\centering
\includegraphics[width=\linewidth]{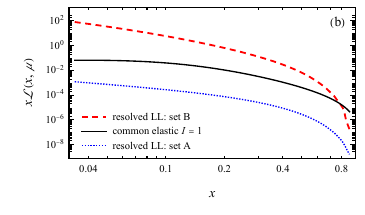}
\end{minipage}
\caption{Elastic saturation and short-distance nonuniqueness.  (a) The cumulative elastic integral $\mathcal L^{\rm el}=\int_{Q_{\min}^2}^{Q_{\rm up}^2}\dd Q^2\,|F(Q^2)|^2/Q^2$ at fixed $x=0.30$, where $Q_{\min}^2=0.11~{\rm GeV}^2$ already exceeds $m_\pi^2$ (cf.\ Eq.~\eqref{eq:xpi}).  A point particle retains the upper-limit logarithm, whereas the pion-pole and lattice-isovector profiles saturate at hadronic virtuality. The separation reaches about three orders of magnitude in the displayed range. The band propagates the lattice covariance within $Q^2\le1~{\rm GeV}^2$, beyond which the fit is not continued.  (b) Pion-pole elastic profile and resolved LL terms at $\mu=2m_t$ for the matched coefficient sets A and B of Eq.~\eqref{eq:degenerate}, all normalized by the same $g_{\rm ref}=|C_3g_AM/f_a|$.  Both sets share the elastic curve. Set A lies below the elastic curve throughout the displayed range; set B lies above it over the dominant rate region and crosses only at large $x$. Table~\ref{tab:rates} gives the process-weighted hierarchy.}
\label{fig:main}
\end{figure*}

Short-distance coefficients and low-energy source directions must also be matched across heavy-quark thresholds. At leading order in the heavy-mass expansion,
\begin{align}
 \widetilde C_G^{(n_f-1)}=\widetilde C_G^{(n_f)}-C_Q,
 \label{eq:threshold}
\end{align}
in the source convention of Eq.~\eqref{eq:qcdsource} \cite{Bauer:2020jbp}. For example, at a five-flavor reference scale take $C_u=-C_d=C_3$ and $C_s=0$.  The two assignments
\begin{align}
 \begin{split}
 {\rm A}:&(C_b,C_c,\widetilde C_G)=(0,0,0),\\
 {\rm B}:&(C_b,C_c,\widetilde C_G)=(C_3,C_3,2C_3)
 \end{split}
 \label{eq:degenerate}
\end{align}
match to the same pure-isovector three-flavor source at this order. They therefore have the same low-energy $g_{aNN}(0)$ and elastic distribution, but different resolved radiation.  Fig.~\ref{fig:main} displays both consequences.

\emph{Numerical results.---}
As an application, we consider an ALP radiated by the proton and a coherent photon radiated by Pb,
\begin{align}
 \sigma_X^{(i)}=&\int_0^1\dd x_a\,\dd x_\gamma\,
 f_{a/p}^{(i)}(x_a,\mu_h)f_{\gamma/{\rm Pb}}(x_\gamma)\nonumber\\
 &\times\hat\sigma_{a\gamma\to X}(\hat s),\qquad
 \hat s=x_ax_\gamma s_{NN},
 \label{eq:totalrate}
\end{align}
with $X=t\bar t,tj$, tree-level hard amplitudes, and $p_T(j)>20$~GeV for $tj$. We use $E_p=7$~TeV and $E_N=2.76$~TeV per nucleon, giving $\sqrt{s_{NN}}\simeq8.8$~TeV.  For resolved radiation the central scale is $\mu_h=\sqrt{\hat s}$. We set $m_a=0$ and consider
ALP emission from the proton accompanied by coherent
photon emission from Pb.

To compare source profiles at fixed couplings, write
\begin{align}
 f^{(i)}_{a/p}(x,\mu)&=\frac{g_{\rm ref}^2}{16\pi^2}x \mathcal{L}_i(x,\mu),\nonumber\\[-2pt]
 \mathcal{L}^{\rm el}(x)&=\int_{Q_{\min}^2(x)}^{Q_{\rm up}^2}
 \frac{\dd Q^2}{Q^2}|F(Q^2)|^2.
 \label{eq:Ldef}
\end{align}
Elastic rows use $g_{\rm ref}=|g_{aNN}(0)|$ and $F=g_{aNN}(Q^2)/g_{aNN}(0)$.  For the lattice profile, the virtuality integral is truncated
at $Q^2=1~{\rm GeV}^2$ and vanishes when $Q_{\min}^2$
reaches or exceeds this limit, while the analytic profiles
are integrated to infinity.  These one-proton-pole integrals describe only the elastic
contribution and should not be identified with the complete
low-scale boundary in Eq.~\eqref{eq:lowboundary}.  For the resolved rows we use Eq.~\eqref{eq:matchpdf} with the coefficient sets of Eq.~\eqref{eq:degenerate} and the common reference $g_{\rm ref}=|C_3g_AM/f_a|$, which equals $|g_{aNN}^{(3)}(0)|$ for both sets.  This normalization permits a scale comparison but does not turn the resolved term into a function of $g_{aNN}(0)$.

Using $\sigma_X[h]$ for Eq.~\eqref{eq:totalrate} with the proton distribution replaced by $h$, define
\begin{align}
 L_X^{(i)}=\frac{\sigma_X[x \mathcal{L}_i(x,\mu_h)]}
 {\sigma_X[x\Theta(x_{\max}-x)]},\qquad
 \rho_X^{(i)}=\frac{L_X^{(i)}}{\ln(E_p^2/M^2)}.
 \label{eq:Leff}
\end{align}
The reference profile $\mathcal{L}_{\rm pt}(x)=\ln(E_p^2/M^2)\Theta(x_{\max}-x)$ gives $17.83$. This $x_{\max}$ restriction applies to the reference and sharp-cutoff rows, while the other numerators retain their own support. Photon-flux normalization cancels in Eq.~\eqref{eq:Leff}, while its shape controls the kinematic weight.

\begin{table}[ht]
	\caption{Process-weighted effective logarithms of Eq.~\eqref{eq:Leff} at $\sqrt{s_{NN}}=8.8$~TeV and $m_a=0$.  The elastic rows are one-proton-pole results normalized by the corresponding $g_{aNN}(0)$, while the lattice interval propagates its covariance within $Q^2\le1~{\rm GeV}^2$, so that its numerator vanishes above $x\simeq0.64$, where $Q_{\min}^2$ leaves the fit range.  The resolved rows are LL components of Eq.~\eqref{eq:matchpdf} for the two coefficient sets of Eq.~\eqref{eq:degenerate}, separately normalized by the common $g_{\rm ref}=|C_3g_AM/f_a|$ and evaluated with $\mu_h=\sqrt{\hat s}$.  The set-B range combines the symmetric 90\% C.L. CT18 Hessian uncertainty with factor-two scale variations in quadrature, while only a central benchmark is quoted for set A, for which light-quark mass and starting-scale uncertainties are not assessed.}
	\label{tab:rates}
	\begin{ruledtabular}
		\begin{tabular}{lcc}
			source profile & $L_{t\bar t}$ & $L_{tj}$\\
			\colrule
			pointlike reference & $17.83$ & $17.83$\\
			sharp $Q<M$ & $0.973$ & $1.10$\\
			\colrule
			\multicolumn{3}{l}{elastic (one-proton pole)}\\
			$I=1$ pion pole & $7.81\times10^{-3}$ & $1.39\times10^{-2}$\\
			$I=1$ lattice & $6.49^{+0.59}_{-0.50}\!\times10^{-3}$ & $1.22^{+0.08}_{-0.07}\!\times10^{-2}$\\
			\colrule
			\multicolumn{3}{l}{resolved quarks (LL component)}\\
			set A: $u+d$ & $5.79\times10^{-5}$ & $1.09\times10^{-4}$\\
			set B: $u+d+c+b$ & $0.81\;(0.66\text{--}0.94)$ & $2.56\;(2.08\text{--}3.03)$\\
		\end{tabular}
	\end{ruledtabular}
\end{table}
Resolved radiation distinguishes the coefficient sets of
Eq.~\eqref{eq:degenerate}: changes in $C_b$ and $C_c$
are compensated by $\widetilde C_G$ in the leading
low-energy source but remain visible in the resolved
quark contribution. With hard couplings fixed,
Table~\ref{tab:rates} gives
$L^{\rm res}_{\rm B}/L^{\rm res}_{\rm A}
=1.4\times10^4$ ($2.4\times10^4$)
for $t\bar t$ ($tj$). Relative to the common lattice
elastic benchmark, the resolved contribution is $0.9\%$
for set A and $125$ ($210$) times as large for set B.
For comparison, after convolution the sharp-cutoff
distribution of Eq.~\eqref{eq:sharp} and the isovector
lattice benchmark suppress the pointlike reference
by factors of $18.3$ ($16.3$) and
$2.7\times10^3$ ($1.5\times10^3$), respectively.

An intact-proton tag isolates the elastic pole, whereas gap- or
veto-based selections can retain low-mass dissociation if the
forward remnants escape detection.  An inclusive selection further admits resolved radiation and requires the matched hard remainder in Eq.~\eqref{eq:subtraction}.  Thus the same low-energy-coupling-normalized flux cannot generically describe these selections.  

\emph{Summary.---}
We have shown that elastic and inclusive ALP
radiation cannot, in general, be described by a common flux
fixed by a single low-energy hadronic coupling.
Elastic ALP radiation is controlled by hadronic
form factors, whereas resolved radiation probes short-distance
quark-flavor couplings. If the pseudoscalar form factor falls as a positive power
of $1/Q^2$ at large virtuality, the elastic distribution
saturates at fixed $x$, eliminating the beam-energy logarithm.
The pion-pole and lattice benchmarks yield a process-weighted
elastic contribution roughly three orders of magnitude below
the pointlike reference. At the same leading low-energy source, two coefficient sets yield resolved LL contributions differing by four orders of magnitude, on opposite sides of the common elastic benchmark. Parity also forbids a local elastic pseudoscalar source for a $0^+$ nucleus. An inclusive prediction requires dissociative input and consistent matching. Resolved radiation thus distinguishes heavy-quark coupling configurations sharing the leading low-energy nucleon projection.

\emph{Acknowledgments.}
This work was supported in part by the National Natural Science Foundation of China under Contract No.~12475098.

\bibliography{axion_compositeness}

\clearpage
\onecolumngrid
\vspace{6pt}
\begin{center}\textbf{End Matter}\end{center}
\vspace{2pt}
\twocolumngrid

\emph{Spectral representation and factorization matching.}
With relativistically normalized states (the sum includes phase-space integration), define
\begin{align}
 W_{IJ}(P,r)={}&\frac12\sum_{s,X}(2\pi)^4\delta^{(4)}(P+r-P_X)
 \nonumber\\
 &\times\langle p(P,s)|O_I^\dagger|X\rangle
 \langle X|O_J|p(P,s)\rangle.
 \label{eq:spectral}
\end{align}
At $O(C^2)$, the equation of motion gives
\begin{align}
 {\cal B}_{a/p}^{\rm cut}(x,\mu_0)={}&\frac{xP^+}{f_a^2}
 \int_{0<Q^2<\mu_0^2}\frac{\dd^4r}{(2\pi)^4}
 \delta(r^++xP^+)\nonumber\\
 &\times\frac{C_I^*W_{IJ}(P,r)C_J}{(Q^2+m_a^2)^2},
 \label{eq:spectralpdf}
\end{align}
where $Q^2=-r^2$, $0<x<1$. Physical final-state support is contained in $W$. With consistently renormalized coefficients and operators, the elastic pole is $2\pi\delta_+((P+r)^2-M^2)Q^2|C_I\mathcal F_I|^2$ in the contracted spectral density. Using $\dd^4r=\dd r^+\dd r^-\dd^2r_T/2$ recovers Eq.~\eqref{eq:elastic}. Independently, the recoil phase space gives $\dd x\,dQ^2/(16\pi^2)$ and the spin trace is $Q^2$. 

Positivity applies to Eq.~\eqref{eq:spectralpdf} and its state-resolved pieces. The $\MSbar$ boundary is
\begin{align}
 \begin{aligned}
 f_{a/p}^{\MSbar}(x,\mu_0)={}&{\cal B}_{a/p}^{\rm el,<}(x,\mu_0)
 +{\cal B}_{a/p}^{\rm diss,<}(x,\mu_0)\\ 
 &+\delta f_a^{\rm cut\to\MSbar}(x,\mu_0),
 \end{aligned}
 \label{eq:conversion}
\end{align}
where the last term converts the explicit virtuality cutoff to the renormalized operator. At a perturbative scale it admits a partonic matching expansion with power corrections, which need not be small at $\mu_0\sim1$~GeV.  

To combine the renormalized ALP distribution with the hard process,
the overlapping collinear contribution must be subtracted~\cite{Collins:1998rz}.
At fixed QCD order the quark subtraction contains
\begin{align}
 \Gamma_{a\leftarrow q}^{\rm LL}(z)=
 z\int_{\mu_q^2}^{\mu^2}\frac{\dd k^2}{k^2}
 \frac{|C_qm_q(k)|^2}{16\pi^2f_a^2}.
 \label{eq:Gamma}
\end{align}
For a hard probe $B$, the matching structure is
\begin{align}
 \dd\sigma_{pB}={}&f_{a/p}^{\MSbar}\otimes \dd\hat\sigma_{aB}^{(0)}
 +\sum_jf_{j/p}\otimes d\hat\sigma_{jB}^{\rm sub},\nonumber\\
 \dd\hat\sigma_{qB}^{\rm sub}={}&\dd\hat\sigma_{qB}^{\rm full}
 -\Gamma_{a\leftarrow q}\otimes \dd\hat\sigma_{aB}^{(0)}.
 \label{eq:subtraction}
\end{align}
The sum includes quarks and gluons. Note that the full $\Gamma$ contains finite terms in the PDF scheme, Eq.~\eqref{eq:Gamma} displays only its logarithm. With QCD evolution, the subtraction must retain the ordered $g\to Q\bar Q\to a$ chain of Eq.~\eqref{eq:matchpdf}. For $f_Q(k)\propto\ln(k^2/m_Q^2)$, ordered integration gives half the double-log coefficient obtained by freezing $f_Q$ at $\mu$. A fixed-scale subtraction cannot be combined with the ordered numerical PDF to claim scale cancellation. The finite $q\gamma\to qX$ remainder, gluonic hard coefficient, and consistent evolution remain required for an inclusive prediction.

\emph{Hard kernels and numerical inputs.}
We use the coherent photon flux of Ref.~\cite{Barbosa:2025zyn}, with $x_\gamma=Ax_A$,
\begin{align}
 f_{\gamma/{\rm Pb}}(x_\gamma)&=\frac{2Z^2\alpha}{\pi x_\gamma}y
 \Big[K_0(y)K_1(y)-\frac y2\big(K_1^2(y)-K_0^2(y)\big)\Big],
 \label{eq:pbflux}
\end{align}
where $y=x_\gamma MR$, $R=7.2~{\rm fm}$, $M$ is the nucleon mass and $R$ is the effective
Pb nuclear radius used in the coherent photon flux. We evaluate the hard kernels for
${\cal L}_{\rm int}=-y_{12}a\,\bar f_1i\gamma_5f_2$,
with the Hermitian conjugate added only for off-diagonal couplings.
For a diagonal coupling, the derivative interaction
$(C_q/2f_a)\partial_\mu a\,\bar q\gamma^\mu\gamma_5q$
gives $y_{qq}=C_qm_q/f_a$, with anomalous gauge terms included
in the source matching. For $a(k_1)\gamma(k_2)\to f_1(p_1)\bar f_2(p_2)$,
\begin{align}
i \mathcal M= i e & Q_fy_{12} \bar u(p_1) \left[
 \slashed\epsilon\frac{\slashed p_1-\slashed k_2+m_1}{(p_1-k_2)^2-m_1^2}i\gamma_5\right.\nonumber\\
 &\left.+i\gamma_5\frac{\slashed k_2-\slashed p_2+m_2}{(k_2-p_2)^2-m_2^2}\slashed\epsilon\right]v(p_2),
 \label{eq:hardamp}
\end{align}
where $y_{tt}=C_tm_t/f_a$. For $tj$ we take a pure axial flavor-changing interaction $(C^A_{tq}/2f_a)\partial_\mu a\,\bar t\gamma^\mu\gamma_5q+\mathrm{h.c.}$, with $q=u,c$. This gives $y_{tq}=C^A_{tq}(m_t+m_q)/(2f_a)$ \cite{Bauer:2020jbp}. General chiral couplings also generate a scalar term. The independent hard couplings $C_t$ and $C^A_{tq}$ are held fixed between sets A and B. We take $m_t=172.5$~GeV and neglect the light-jet mass. We check the amplitude using the Ward identity and independent spinor/trace evaluations, and the convolution using the constant reference and Eq.~\eqref{eq:closed}.

\emph{Elastic source profiles.}
We now specify the elastic source profiles entering the numerical convolution. All source directions in Eq.~\eqref{eq:sourcebasis} are defined at a common renormalization scale and in a common scheme; throughout, we write $C_G\equiv\widetilde C_G$. Here $0$ labels the light-quark isoscalar direction, while the independent strange direction is omitted.

The lattice input is $G_5$, related to the axial and induced pseudoscalar form factors by $\hat mG_5/M=G_A-Q^2G_P/(4M^2)$. Its normalized form is
\begin{align}
 F_3(Q^2)&=\frac{G_5(Q^2)}{G_5(0)}=\frac{m_\pi^2}{m_\pi^2+Q^2}
 \frac{\sum_{k=0}^3a_kz(Q^2)^k}{\sum_{k=0}^3a_kz(0)^k},\nonumber\\
 z(Q^2)&=\frac{\sqrt{t_{\rm cut}+Q^2}-\sqrt{t_{\rm cut}+t_0}}
 {\sqrt{t_{\rm cut}+Q^2}+\sqrt{t_{\rm cut}+t_0}},
 \label{eq:isovector}
\end{align}
We use $m_\pi=0.135$~GeV and $t_{\rm cut}=9m_\pi^2$. Here $t_0$ denotes the reference value of $Q^2$, with $z(t_0)=0$. The $G_5$ fit in Appendix A.3 of Ref.~\cite{Alexandrou:2023qbg} uses $t_0=-m_\pi^2$; its $G_A$ fit uses $t_0=0$. The quoted $G_5$ coefficients are $\bm a=(4.62,-2.2,-2.9,-1.2)$, $\bm\sigma=(0.33,2.5,3.7,2.4)$, and
\begin{align}
 \rho=\begin{pmatrix}
 1&-0.804&0.435&0.140\\
 -0.804&1&-0.825&0.217\\
 0.435&-0.825&1&-0.694\\
 0.140&0.217&-0.694&1
 \end{pmatrix}.
 \label{eq:cov}
\end{align}
We quote the central $68\%$ intervals, defined by the
16th and 84th percentiles of the propagated distributions.
The $16$--$84\%$ intervals propagate $2\times10^5$ Gaussian draws with $V_{ij}=\sigma_i\rho_{ij}\sigma_j$, renormalizing $F_3(0)=1$ for each draw. Extending the fit with its central coefficients beyond
$Q^2=1~{\rm GeV}^2$ increases the truncated integrals by
$0.9\%$ ($t\bar t$) and $0.5\%$ ($tj$).
This diagnostic does not constrain the true high-$Q^2$ tail,
whose uncertainty is not included in the quoted intervals.

Beyond the lattice-constrained isovector channel, we use
simple pole-based profiles to illustrate the possible
momentum dependence of the isoscalar and gluonic sources: 
\begin{align}
 F_0(Q^2)&=\left[(1-w)\frac{m_\eta^2}{m_\eta^2+Q^2}
 +w\frac{m_{\eta'}^2}{m_{\eta'}^2+Q^2}\right]\nonumber\\
 &\quad\times\left(1+\frac{Q^2}{\Lambda_0^2}\right)^{-2},
 \label{eq:isoenv}\\
 F_G(Q^2)&=\frac{m_{\eta'}^2}{m_{\eta'}^2+Q^2}
 \left(1+\frac{Q^2}{\Lambda_G^2}\right)^{-1},
 \label{eq:genv}
\end{align}
with $0\le w\le1$, $1.0\le\Lambda_0/{\rm GeV}\le1.4$, and $1.0\le\Lambda_G/{\rm GeV}\le2.0$.  Continuing the central $F_3$ fit for this illustration, $F=(F_3+\eta_0 F_0)/(1+\eta_0)$ gives $L_{t\bar t}=0.037$--$0.089$ for $\eta_0=1$ and $0.003$--$0.024$ for $\eta_0=-0.3$, where $\eta_0\equiv g_0/g_3$. At $\eta_0=-1$ the normalization by $g_{aNN}(0)$ fails.

\begin{figure}[t]
\centering
\includegraphics[width=0.98\columnwidth]{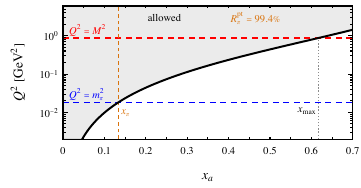}
\caption{Elastic-emission kinematics for the $p$Pb $t\bar t$ example. Emission is allowed above $Q_{\min}^2(x_a)$ (grey region). The orange line marks $x_\pi=0.134$, where $Q_{\min}^2=m_\pi^2$. For the normalized pointlike partonic weight restricted to $x_a<x_{\max}$,
$R_\pi^{\rm pt}\equiv\sigma_{\rm pt}(x_\pi<x_a<x_{\max})/\sigma_{\rm pt}(x_a<x_{\max})=0.994$.}
\label{fig:kin}
\end{figure}

\emph{Heavy-final-state support.}
To relate the elastic-saturation theorem to the heavy-final-state benchmarks in the Letter, we ask when the kinematic lower limit lies below the pion scale.
From Eq.~\eqref{eq:kin}, $Q_{\min}^2<m_\pi^2$ requires $x_a<x_\pi$, with
\begin{align}
 x_\pi=\frac{\sqrt{r_\pi^2+4r_\pi}-r_\pi}{2}\simeq0.134,
 \qquad r_\pi=\frac{m_\pi^2}{M^2}.
 \label{eq:xpi}
\end{align}
This is necessary, but not sufficient, for $Q^2<m_\pi^2$. Fig.~\ref{fig:kin} shows the correlation. At $\sqrt{s_{NN}}=8.8$~TeV, the $t\bar t$ production
threshold requires
$y\ge y_{\min}=4m_t^2MR/(x_as_{NN})\simeq0.053/x_a$.
Smaller $x_a$ therefore probes larger $y$, where the
photon flux in Eq.~\eqref{eq:pbflux} is suppressed.

Convolving the photon flux in Eq.~\eqref{eq:pbflux} with
$\hat\sigma_{a\gamma\to t\bar t}\propto
\operatorname{artanh}\beta/\hat s$,
where $\beta=\sqrt{1-4m_t^2/\hat s}$, gives a fraction
of only $0.6\%$ below $x_\pi$ for the pointlike reference
of Ref.~\cite{Barbosa:2025zyn}, restricted to
$x_a<x_{\max}$. Equivalently, $R_\pi^{\rm pt}\equiv
\sigma_{\rm pt}(x_\pi<x_a<x_{\max})
/\sigma_{\rm pt}(0<x_a<x_{\max})=0.994$;
the corresponding value for $tj$ is $0.986$.
The fractions below $x_\pi$ increase to $2.7\%$ and
$30\%$ for the sharp-cutoff and pion-pole profiles,
respectively, compared with $5.9\%$ and $48\%$ for $tj$.

The corresponding integrated rates can be compared
directly using Table~\ref{tab:rates}.
At fixed couplings, Table~\ref{tab:rates} gives
$\sigma_{Q<M}/\sigma_{\rm pt}\simeq0.0546$ and
$\sigma_\pi/\sigma_{Q<M}\simeq0.0080$, yielding
$\sigma_\pi/\sigma_{\rm pt}\simeq4.4\times10^{-4}$
for the pure-isovector $t\bar t$ pion-pole benchmark.

The pion-pole numerator is integrated over its full
support, $0<x_a<1$. Restricting it to the reference
interval $x_a<x_{\max}$ changes $L_{t\bar t}$ by only
$1.2\%$, from $0.007809$ to $0.007716$.
All quoted fractions are evaluated at parton level,
before detector selection.

\emph{Dissociative kernel illustration.}
To isolate the kinematic effect of a heavier hadronic final
state, we replace only the lower virtuality limit in
Eq.~\eqref{eq:closed}, keeping the elastic pion-pole kernel
otherwise unchanged. The ratio $S$ of the resulting
process-weighted integral to the elastic one is $S=0.218,0.078,0.024,0.006$ at $M_X=M+m_\pi,1.232,1.5,2.0$~GeV in $t\bar t$ ($0.198,0.068,0.021,0.005$ in $tj$). This illustrative kernel falls as $M_X^{-4}$ at large $M_X$. Physical dissociation also changes the transition
matrix element. For a same-parity spin-$1/2$ transition, the spin-averaged
numerator becomes $Q^2+(M_X-M)^2$, illustrating the additional
transition dependence beyond the elastic pole.
Other final states require their own spin structures and,
for multiparticle states, phase-space integrals, together
with transition form factors and non-pole contributions.
The quoted ratios thus quantify the kinematic suppression
from the raised threshold. A dissociation rate additionally
requires the corresponding transition spectral strength.

\emph{Heavy-flavor matching and resolved numerical inputs.}
Neglecting electroweak interactions, we write the axial anomaly~\cite{Adler:1969gk,Bell:1969ts} in the convention
$\partial_\mu(\bar Q\gamma^\mu\gamma_5Q)=2m_Q\bar Qi\gamma_5Q+2\Qtop$.
Decoupling the heavy current gives $m_Q\bar Qi\gamma_5Q=-\Qtop+O(\mathcal O_6/m_Q^2)$ in low-energy matrix elements~\cite{Franz:2000ee}, yielding Eq.~\eqref{eq:threshold}. Applying it to Eq.~\eqref{eq:degenerate}, set B runs $\widetilde C_G=2C_3\to C_3\to0$ across the $b$ and $c$ thresholds and set A stays at $\widetilde C_G=0$, so both reach the same three-flavor source $(C_u,C_d,C_s,\widetilde C_G)=(C_3,-C_3,0,0)$. This degeneracy holds at leading order in running and the heavy-mass expansion, independently of the fixed hard couplings $C_t$ and $C^A_{tq}$.

With the coefficient sets matched to the same low-energy source,
we evaluate Eq.~\eqref{eq:matchpdf} for Table~\ref{tab:rates}
using CT18NNLO PDFs~\cite{Hou:2019efy}
and the associated $\as$, with one-loop running quark masses,
$\dd\ln m_q/\dd\ln\mu^2=-\as/\pi$.
The mass inputs are
$m_u(2~{\rm GeV})=2.16$~MeV,
$m_d(2~{\rm GeV})=4.67$~MeV~\cite{ParticleDataGroup:2022pth},
$m_c(m_c)=1.275$~GeV, and
$m_b(m_b)=4.18$~GeV~\cite{ParticleDataGroup:2022pth}.  The charm and bottom activation thresholds are $1.30$ and $4.75$~GeV, and for light quarks, the LL evolution starts at
$\mu_0=1.295$~GeV.  Set A receives only the $u+d$ term of Eq.~\eqref{eq:matchpdf}; set B adds $c$ and $b$, while its $\widetilde C_G$ generates no LL kernel.  Fig.~\ref{fig:main}(b) evaluates both at the fixed scale $2m_t$ and Table~\ref{tab:rates} at $\mu_h=\sqrt{\hat s}$ inside the convolution.  Bottom-quark radiation accounts for $83\%$ of the set-B
resolved LL contribution, owing to the quark-mass-squared
weighting in Eq.~\eqref{eq:matchpdf}.  Only a central benchmark is quoted for set A; light-quark mass and starting-scale uncertainties are not assessed.
The ranges quoted for set B in Table~\ref{tab:rates}
combine the symmetric 90\% C.L. CT18 Hessian uncertainty,
$\Delta L_{\rm PDF}=\frac12[\sum_j(L_j^+-L_j^-)^2]^{1/2}$,
in quadrature with the envelope of factor-two scale variations.
To quantify the effect of scale ordering, we repeat the
convolution with the heavy-quark PDFs evaluated at
$\mu_h$ throughout. This gives $L_{t\bar t}=1.12$ and
$L_{tj}=3.94$, compared with $0.81$ and $2.56$ from
the ordered integration.

\end{document}